\documentclass{elsart}
\usepackage{amssymb}
\usepackage{graphicx}

\journal{Physica A}

\begin{document}

\begin{frontmatter}

\title{First-order transition features of the triangular Ising model with nearest- and next-nearest-neighbor
antiferromagnetic interactions}

\author{A. Malakis\corauthref{cor1}}, \corauth[cor1]{Corresponding author.}
\ead{amalakis@phys.uoa.gr} \author{N. G. Fytas} and
\author{P. Kalozoumis}

\address{Department of Physics, Section of Solid State Physics, University of Athens, Panepistimiopolis, GR 15784
Zografos, Athens, Greece}

\begin{abstract}
We implement a new and accurate numerical entropic scheme to
investigate the first-order transition features of the triangular
Ising model with nearest-neighbor ($J_{nn}$) and
next-nearest-neighbor ($J_{nnn}$) antiferromagnetic interactions
in ratio $R=J_{nn}/J_{nnn}=1$. Important aspects of the existing
theories of first-order transitions are briefly reviewed, tested
on this model, and compared with previous work on the Potts model.
Using lattices with linear sizes
$L=30,40,\ldots,100,120,140,160,200,240,360$ and $480$ we estimate
the thermal characteristics of the present weak first-order
transition. Our results improve the original estimates of Rastelli
et al. and verify all the generally accepted predictions of the
finite-size scaling theory of first-order transitions, including
transition point shifts, thermal, and magnetic anomalies. However,
two of our findings are not compatible with current
phenomenological expectations. The behavior of transition points,
derived from the number-of-phases parameter, is not in accordance
with the theoretically conjectured exponentially small shift
behavior and the well-known double Gaussian approximation does not
correctly describe higher correction terms of the energy
cumulants. It is argued that this discrepancy has its origin in
the commonly neglected contributions from domain wall corrections.
\end{abstract}

\date{\today}


\end{frontmatter}

\section{Introduction}
\label{sec:1}

Generalizations of the Ising model including further than
nearest-neighbor interactions may give rise to complicated spatial
orderings and produce complex and rich critical
behavior~\cite{selke92,lawrie92}. The transitions between ordered
and disordered phases may be continuous or of first-order with a
tricritical point between them. The exactly soluble Ising model in
$d=2$~\cite{onsager} with the addition of next-nearest-neighbor
interactions becomes an analytically intractable problem and an
understanding of the effects of adding such further couplings on
the critical behavior of the system is an open important problem
of great theoretical and experimental interest. We will be
particularly interested is cases involving competing interactions
with ground-state arrangements that mimic a sublattice order or
superantiferromagnetic (SAF) order in which ferromagnetic lines
along the directions of the lattice alternate with lines of
opposite oriented spins. Such models are of great theoretical and
experimental interest. They simulate various types of
antiferromagnets~\cite{binder87,tanaka75,phani80,oitmaa81,velgakis88},
but also important models of alloy orderings~\cite{binder87}. Due
to the well-known correspondence between spin systems and lattice
gas, they provide an understanding of the behavior of adsorbed
gases on crystal surfaces~\cite{landau83,bretz77}. It should be
noted here that, remarkable efforts have been made to predict the
order of the transition according to the Landau-Lifshitz rules,
which, on the basis of group theory arguments, select the ordered
configurations that can be reached from the disordered phase by a
continuous transition~\cite{binder87,domany78}. K.
Binder~\cite{binder87} has reviewed these rules and has also
pointed out well-known counter-examples, emphasizing the
experimental difficulties in distinguishing a weak first-order
transition from a second-order transition.

Second-order transitions are more special than first-order
transitions, but they are theoretically much better understood. At
a first-order transition there is no diverging correlation length,
and in general one cannot restrict attention to long wavelength
phenomena, thus no universality, as in critical phenomena, is to
be expected. Furthermore, it is well known that antiferromagnetic
arrangements, with several order-parameter components, may be
produced by the competition of the interactions~\cite{binder87}.
However, no general theoretical agreement exists connecting the
symmetry of spin structures, the number of order-parameter
components, and the range of interaction with the expected
critical behavior and in particular the order of the phase
transition. Furthermore, many authors have demonstrated the
difficulties in properly identifying the order of the transition
on a ground of high-temperature expansion, scaling,
renormalization group transformations, and Monte Carlo simulations
(see Ref.~\cite{binder87} and references therein). Roomany and
Wyld~\cite{roomany81} pointed out that the occurrence of
second-order type effects at the weakly first-order transitions
can be explained by a comparison of the correlation length $\xi$
with the lattice size $L$. In the case of the $q=5$ Potts model's
weak first-order transition, Peczak and Landau~\cite{peczak89}
have observed pseudocritical exponents close to the conjectured
values of the critical indices in the $q=4$ Potts model.

This paper is concerned with the analysis of numerical data,
obtained via an accurate entropic Monte Carlo scheme, on
triangular Ising finite systems with nearest-neighbor ($J_{nn}$)
and next-nearest-neighbor ($J_{nnn}$) antiferromagnetic
interactions in ratio $R=J_{nn}/J_{nnn}=1$. Rastelli et
al.~\cite{rastelli05} have recently presented numerical evidence
(for $R=0.1$, $0.5$, and $1.0$) that, in the thermodynamic limit,
this model undergoes a first-order phase transition, from a
layered ordered phase (SAF phase) to a high temperature
paramagnetic phase. Here, we will present a more detailed
identification of the nature of this transition by looking again
at the size dependence of the traditional thermodynamic quantities
but also by implementing the Lee-Kostelirlitz
method~\cite{lee90,lee91} and detecting the first-order character
of the transition via the size-dependence of the free energy
barrier. Furthermore, we will improve the original estimates of
Rastelli et al.~\cite{rastelli05} and try to verify the
predictions of the finite-size scaling (FSS) theory of first-order
transitions, including transition point shifts, thermal, and
magnetic anomalies.

The theory of FSS near second-order transitions has a rich and
longstanding literature~\cite{barber83}, starting with the
pioneering works of Fisher~\cite{fisher71} and Fisher and
Barber~\cite{fisher72}. This theory has been extended to
first-order
transitions~\cite{binder87,lee90,lee91,binder84,challa86,privman90,binder92}
and FSS and renormalization group methods have proven to be very
useful tools for the study of first- and second-order phase
transitions~\cite{lee90,lee91,barber83,fisher71,fisher72,binder84,challa86,privman90,binder92,fisher82,nienhuis79}.
Our attempt here to elucidate the distinctive first-order features
of the present model, by an extensive numerical study, will
closely follow previous analogous studies carried out on the
$q=5$, $q=8$, and $q=10$ Potts
model~\cite{lee90,lee91,challa86,borgs92a,janke93}. In these
studies the existing theories of first-order transitions have been
tested and verified but several important aspects have not been
thoroughly clarified, especially in the cases of weak first-order
transitions. Not only the demonstration of a weak first-order
transition is more difficult but also strong finite-size effects
may obscure the true asymptotic behavior~\cite{lee91} and some
theoretical predictions may not be met even at quite large
lattices (since the correlation length may be very large). In such
cases, it is very difficult to discriminate between wrong
phenomenological expectations and correct theoretical predictions.
Therefore, comparisons of different models and alternative studies
may provide useful information and a better understanding of
finite-size effects, something that is necessary for the correct
interpretation of the numerical data and the verification of the
theories. The present model, showing a weak first-order phase
transition, offers the opportunity of such a contrasting test with
the well-studied cases of the Potts model.

Our interest in the present model stems from our recent studies on
the analogous square SAF model~\cite{malakis06,malakis07}, which
is again an Ising model on the square lattice with
nearest-neighbor ($J_{nn}$) and next-nearest-neighbor ($J_{nnn}$)
antiferromagnetic interactions. The ground state of the square SAF
model is four-fold degenerate and consists of the arrangements in
which ferromagnetic rows (columns) alternate with opposite
oriented spins. In the square model the SAF order can be obtained
in both cases of a ferromagnetic or an antiferromagnetic
nearest-neighbor coupling (for its $T=0$ phase diagram see
Refs.~\cite{oitmaa81,landau85,minami94}). Similarly, the present
triangular SAF model, with antiferromagnetic interactions in ratio
$R=J_{nn}/J_{nnn}=1$, has a six-fold degenerate ground state and
consists of the six arrangements in which ferromagnetic lines
alternate with opposite oriented spins in the three lattice
directions ($T=0$ phase diagrams are given in
Refs.~\cite{tanaka75,landau83}). The Hamiltonian that governs
these systems, in zero-field, is
\begin{equation}
\label{eq:1} \mathcal{H}=J_{nn}\sum_{\langle
i,j\rangle}\sigma_{i}\sigma_{j}+J_{nnn}\sum_{(i,j)}\sigma_{i}\sigma_{j},
\end{equation}
where here both nearest-neighbor ($J_{nn}$) and
next-nearest-neighbor ($J_{nnn}$) interactions will be assumed to
be positive (antiferromagnetic). This Hamiltonian has been studied
also in $d=3$ FCC lattices and evidence for both first- and
second-order phase transitions have been
presented~\cite{binder87,phani80}, although the distinction of the
order of the transition was difficult in that case too.

The square model, governed by Eq.~(\ref{eq:1}), develops at low
temperatures SAF order for $R>0.5$ and several early
studies~\cite{oitmaa81,landau85,minami94,nightingale77,swendsen79,binder80,landau80},
mainly based on importance
sampling~\cite{metropolis53,bortz75,binder97,newman99,landau00},
have suggested that this system possess anomalous exponents and a
non-universal critical behavior with exponents depending on the
coupling ratio $R$. However, despite this early accepted scenario,
there has been recently renewed interest and some attempts to
re-examine the behavior of this model appeared. In several papers
Lopez et al.~\cite{lopez93,lopez94,lopez99} have used the cluster
variation method (CVM) to study this model, concluding that the
system undergoes a first-order transition for a particular range
of the coupling ratio $R$ ($R=0.5-1.2$). On the other hand, this
different scenario - predicting first-order transitions between
ordered and disordered phases followed by continuous transitions
outside the first-order region - has now been seriously questioned
by the present authors, who have presented very strong
evidence~\cite{malakis06} for the case $R=1$ that points out a
well-obeyed second-order scaling behavior. The Buzano and
Pretti~\cite{buzano97} attempt to verify the scenario of Lopez et
al.~\cite{lopez93,lopez94,lopez99}, by including an additional
$4$-body coupling, and using again the CVM, has revealed a further
peculiar inadequacy of the CVM. The limiting case ($J_{nn}=0$),
where the exact solution of the Baxter model~\cite{baxter72}
applies, was also predicted by the CVM to undergo a first-order
transition, in contradiction to the exact solution.

The above introduction describes a rather controversial situation
on the class of models with antiferromagnetic ground-state
arrangements. Therefore, it is of interest to follow closely the
previous FSS analysis applied to the Potts models and study in
more detail the triangular SAF model. In this case, we will insist
that this system undergoes a genuine, but weak, first-order
transition, as originally predicted by Rastelli et
al.~\cite{rastelli05}. Having a six-fold degenerate ground state
arrangement, this model will be shown to produce first-order
characteristics that lie between the $q=5$ and $q=6$ Potts model.
The rest of the paper is organized as follows. In the next Section
we outline an extensive entropic sampling program. This program
goes beyond our previous practice in other
applications~\cite{malakis04,malakis05,martinos05,fytas06} and is
based on (i) the Wang-Landau (WL) method~\cite{wang01}, (ii) our
dominant energy restriction (CrMES) scheme~\cite{malakis04}, and
(iii) a second-stage improvement that combines the
WL~\cite{wang01}, the Lee entropic~\cite{lee93,lee06}, and the
broad histogram (BH)~\cite{oliveira96} or transition
matrix~\cite{wang99} methods. In Section~\ref{sec:3} we shall
review all necessary theoretical aspects of first-order
transitions that are then used for the analysis of our numerical
data. The free energy barrier method of Lee and
Kosterlitz~\cite{lee90,lee91}, used in the literature for the
identification of a first-order transition, is discussed in
subsection~\ref{sec:3aa}, while the size-dependence of thermal
properties derived from the well-known double Gaussian
approximation is summarized in subsection~\ref{sec:3b}. Finally,
some new transition points derived from the number-of-phases
parameter~\cite{borgs92a,janke93} are presented in
subsection~\ref{sec:3c}. Section~\ref{sec:4} presents the analysis
of our numerical data. In subsection~\ref{sec:4a} we make use of
several alternative methods for the estimation of the transition
temperature presenting a detailed analysis of the new transition
points derived from the number-of-phases parameter and a
discussion on their conjectured exponentially small shift
behavior. In subsection~\ref{sec:4b} we explore our data for the
energy cumulants and the magnetic susceptibility, emphasizing on
the examination of the higher-order corrections predicted by the
double Gaussian approximation. The characteristic values of the
energy cumulants, i.e. the coefficient of the dominant diverging
term of the specific heat and the limiting values of the second-
and fourth-order energy cumulants are determined and the
corresponding theoretical predictions are critically discussed.
This serves as a useful self-consistency test of our numerical
data but also of the theoretical predictions. Finally, our
conclusions are summarized in Section~\ref{sec:5}.

\section{Numerical approach}
\label{sec:2}

Computer simulations based on Monte Carlo sampling methods have
increased dramatically our understanding of the behavior of
systems of classical statistical mechanics. The Metropolis method
and its variants were for many years the main tools in condensed
matter physics and critical
phenomena~\cite{metropolis53,bortz75,binder97,newman99,landau00}.
However, for complex systems, effective potentials may have
competing minima, or a rugged landscape, becoming more pronounced
with increasing system size. In such cases, the traditional
methods become completely inefficient, since they cannot overcome
such barriers and cross from one basin to another in the state
space. On the other hand, a large number of ``generalized
ensemble'' methods have been proposed to overcome such
troubles~\cite{newman99,landau00,wang01,lee93,lee06,oliveira96,wang99,berg92,smith95,torrie97,swendsen86,geyer91,marinari92,lyubartsev92,hukushima96,marinari98}.
One important class of these methods emphasizes the idea of
directly sampling the energy density of states (DOS) and may be
called entropic sampling methods~\cite{newman99}. In an enropic
sampling method, instead of sampling microstates with probability
proportional to $e^{-\beta E}$, we sample microstates with
probability proportional to $[G(E)]^{-1}$, where $G(E)$ is the
DOS, thus producing a ``flat energy histogram''. The prerequisite
for the implementation of the method, is the DOS information of
the system, a problem that can now be handled in many adequate
ways. Over the last two decades, there have been a number of
interesting approaches addressing this problem. The most
remarkable examples are the Lee entropic~\cite{lee93,lee06}, the
multicanonical~\cite{berg92,smith95}, the BH~\cite{oliveira96},
the transition matrix~\cite{wang99}, and the WL~\cite{wang01}
methods. In particular, the WL algorithm~\cite{wang01} has been
one of the most refreshing improvements in Monte Carlo simulation
schemes and has been already applied to a broad spectrum of
interesting problems in statistical mechanics and
biophysics~\cite{yamaguchi01}.

Recently, the present authors have
introduced~\cite{malakis04,malakis05} a dominant energy subspace
implementation of the above entropic methods, called critical
minimum energy subspace (CrMES) method. The main idea is a
systematic restriction of the energy space with increasing lattice
size. The (WL) random walk takes place in the appropriately
restricted energy space and this restriction produces an immense
speed up without introducing observable errors. For the
temperature range of interest, that is the range around a critical
or, for the present model, a first-order transition point, this
scheme may be used to determine all finite-size thermal anomalies
from the final accurate DOS of the WL scheme. A further
simplification, for the determination of  the magnetic finite-size
anomalies, was suggested~\cite{malakis05} by using the WL approach
as a one-run entropic strategy. In this CrMES WL-entropic sampling
method the magnetic properties are obtained by recording
appropriate histograms in the high levels (or final levels) of the
WL process. It was argued and numerically verified that these high
levels, in which the detailed balanced condition is quite well
satisfied, give good approximations for the microcanonical
estimators, necessary for the evaluation of other properties,
besides the thermal ones, of the statistical system. The method
was efficiently combined with the N-fold
way~\cite{bortz75,malakis05,schulz02,malakis04b} in order to
improve statistical reliability, but also to produce BH
estimators, from the same high levels, for an additional
calculation of the DOS.

Let us now outline the main ingredients and equations used in the
above described implementation. The approximation of canonical
averages, in a temperature range of interest, is as follows:
\begin{equation}
\label{eq:2} \langle Q\rangle=\frac{\sum_{E}\left<
Q\right>_{E}G(E)e^{-\beta E}}{\sum_{E}G(E)e^{-\beta E}}\cong
\frac{\sum_{E\in(E_{1},E_{2})}\langle
Q\rangle_{E,WL}\widetilde{G}(E)e^{-\beta
E}}{\sum_{E\in(E_{1},E_{2})}\widetilde{G}(E)e^{-\beta E}}.
\end{equation}
The restricted energy subspace $(E_{1},E_{2})$  is carefully
chosen to cover the temperature range of interest without
introducing observable errors. The microcanonical averages
$\langle Q\rangle_{E}$ are determined from the $(E,Q)$-histograms
(denoted by $H_{WL}(E,Q)$), which are obtained during the high
levels of the WL process
\begin{equation}
\label{eq:3} \langle Q\rangle_{E}\cong\langle
Q\rangle_{E,WL}\equiv \sum_{Q}Q\frac{H_{WL}(E,Q)}{H_{WL}(E)};\;\;
H_{WL}(E)=\sum_{Q}H_{WL}(E,Q)
\end{equation}
and the summations run over all values generated in the restricted
energy subspace $(E_{1},E_{2})$. Finally, the approximate DOS used
in Eq.~(\ref{eq:2}) is the final ``most accurate'' DOS generated
via the WL iteration $[\widetilde{G}(E)=G_{WL}(E)]$ or
alternatively from the accumulated BH approximation
$[\widetilde{G}(E)=G_{BH}(E)]$. The initial modification factor of
the WL process is taken to be $f_{1}=e=2.718\ldots$ and, as usual,
we follow the rule $f_{j+1}=\sqrt{f_{j}}$ and a $5\%$ flatness
criterion~\cite{malakis04,malakis05}. Details on the N-fold
implementation can be found in
Refs.~\cite{malakis05,fytas06,schulz02,malakis04b}. As mentioned
above, the recording of the appropriate histograms takes place in
the high levels ($j=18-26$) of the WL process, where the
incomplete detailed balance condition has no significant effects.

The CrMES restriction may be defined by requesting a specified
accuracy on a diverging specific heat (and/or on a diverging
susceptibility)~\cite{malakis05}. Alternatively, the energy
density function may be used to restrict the energy space. This
latter restriction is very simple for presentation reasons (see
below) but we always use the original recursion method described
in Ref.~\cite{malakis04}, which from our experience produces the
most accurate and safe location of the dominant energy subspace.
Consider a temperature of interest $T^{*}_{L}$ and let
$\widetilde{E}$ be the value maximizing the probability density,
at this temperature. The end-points $(\widetilde{E}_{\pm})$ of the
dominant energy subspaces may be located by the above mentioned
simple energy density function condition
\begin{equation}
\label{eq:4} \widetilde{E}_{\pm}:
\frac{P_{\widetilde{E}_{\pm}}(T_{L}^{*})}{P_{\widetilde{E}}(T_{L}^{*})}\leq
r,
\end{equation}
where $r$ is chosen to be a small number, independent of the
lattice size (i.e. $r=10^{-6}$). Similarly, let $\widetilde{M}$ be
the value maximizing the order-parameter density at some
pseudocritical temperature, for instance the pseudocritical
temperature of the susceptibility
$(T_{L}^{*}=T_{L,\chi_{max}}^{*})$. Then, the end-points
$(\widetilde{M}_{\pm})$ of the dominant magnetic subspaces may be
located by a similar condition
\begin{equation}
\label{eq:5} \widetilde{M}_{\pm}:
\frac{P_{\widetilde{M}_{\pm}}(T_{L}^{*})}{P_{\widetilde{M}}(T_{L}^{*})}\leq
r.
\end{equation}
The finite-size extensions of the above dominant energy and
order-parameter subspaces satisfy respectively the scaling laws of
the specific heat and susceptibility with exponents $\alpha/\nu$
and $\gamma/\nu$,
respectively~\cite{malakis06,malakis07,malakis04,malakis05}.

The performance limitations of entropic methods, such as the WL
random walk and the reduction of their statistical fluctuations
have recently attracted considerable
interest~\cite{lee06,dayal04,zhou05}. For the CrMES entropic
scheme, presented above, comparative studies using various
implementations and the Metropolis algorithm were presented in
Refs.~\cite{malakis04,malakis05} for the $d=2$ and $d=3$ Ising
models and also in Refs.~\cite{malakis06,malakis07} for the square
SAF model. In particular, the effects of the used range of the WL
iteration levels on the magnetic properties of the system and also
the effect of some refinements of the WL algorithm were discussed.
However, for large systems statistical fluctuations significantly
increase and their reduction demands, as usually, multiple
measurements, i.e. carrying out many WL random walks. The
recordings of the $(E,Q)$-histograms in the high levels of the WL
iteration process will in general produce quite accurate estimates
for the magnetic or other properties, but it should be noted that
this accuracy may well depend on the lattice sizes used but also
on the particular model simulated. From our experience, all
ingredients of the entropic schemes should be tested and treated
with caution. There are, for instance, several ideas in the
literature for reducing the modification factor of the WL process
but also for the proper use, i.e. recording enough statistics, of
the histogram-flatness criterion of the WL
method~\cite{malakis05,wang01,lee06,zhou05}.

\begin{figure}[htbp]
\includegraphics{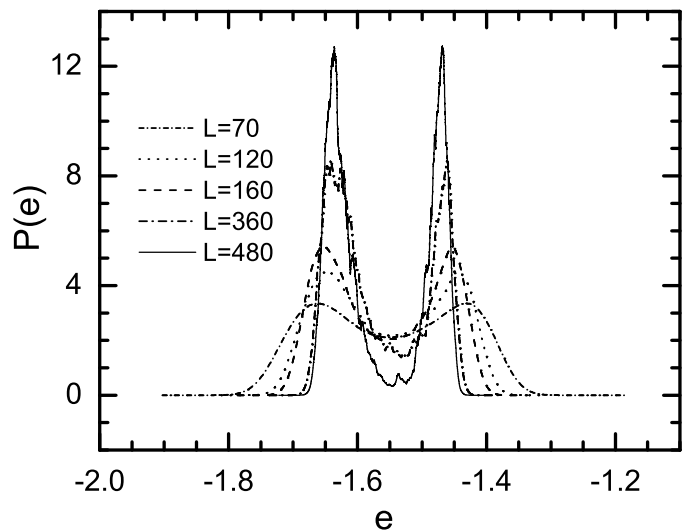}
\caption{\label{fig:1}Energy distributions at $T_{h}$ for the
$R=1$ triangular SAF model for selected lattice sizes.}
\end{figure}

Taking all these into account, we first applied a
multi-energy-range approach~\cite{wang01} using the one-run
entropic scheme of Ref.~\cite{malakis05}, recording the
$(E,M)$-histograms only in the final levels $j=18-26$ of the WL
process. For all lattice sizes
($L=30,40,\ldots,100,120,140,160,200,240,360$ and $480$) this
simulation was repeated for $10$ independent random walks and
averages of the resulting DOS and histograms were constructed.
These time-demanding simulations have been greatly facilitated by
the dominant energy subspace restriction applied. Even by using an
$r=10^{-6}$ accuracy level, we had to simulate only a small part
of the order of $4-5\%$ of the energy spectrum for the larger
lattices. For $L=240$ we determined the DOS for about $5400$
energy levels ($4.7\%$), which is about $10\%$ wider than the
resulting dominant subspace of $5000$ energy levels. For $L=480$
we carried out the simulation in a subspace of $19000$ energy
levels ($4\%$) and the resulting dominant subspace was about
$15500$ energy levels ($3.36\%$). These numerical data were used
to observe the first-order nature of the present model and in fact
the resulting evidence was decisively supporting the first-order
character of the transition (note that Figs.~\ref{fig:1} and
\ref{fig:2} in the next Section are using also these data).
However, due to the first-order character the statistical
fluctuations for the present model were rather large, compared to
other simpler models studied earlier by the same
scheme~\cite{malakis06,malakis04,malakis05,martinos05}.
Accordingly we have carried out a second-stage entropic sampling
scheme only for the lattices $L=40,\ldots,240$. In this scheme we
performed two parallel random walks, one using the $j=26$
modification factor of the WL method (WL-walk) and the other using
the Lee correction~\cite{newman99,lee93,lee06} of the DOS
(Lee-walk). The total duration of this process was set equal to
$10\times \tau_{26}$, where $\tau_{26}$ is the time needed in the
original scheme for the saturation of histogram
fluctuations~\cite{lee06} of the WL scheme at the level $j=26$.
Details of these additional simulations are given below.

The second-stage simulation was repeated three times for each
lattice size, every time starting with a different DOS, selected
from our first one-run approach. In time intervals equal to the
time-unit $\tau_{26}$, the resulting DOS's from the two parallel
walks (WL-walk and Lee-walk) were recorded, as were also recorded
the corresponding ``microcanonical estimators'' for the
application of the BH or transition matrix methods. To continue in
the next time-step (of duration $\tau_{26}$) we used for both
parallel walks the time-accumulated average DOS of the Lee-walk.
Of course, this practice is one of many other possibilities and
ideas for ``refreshing'' the DOS, that could be used to continue
the further refinement of the DOS via a WL-modification ($j=26$)
or via a Lee-adjustment~\cite{lee93,lee06}. One can even mix the
DOS's of the two parallel walks and/or combine with the DOS's
resulting from the BH or transition matrix methods to continue the
process (such an example was considered by Shell et al. in
Ref.~\cite{yamaguchi01}). At the end of the process, we have
observed the time development of various thermal properties of the
system at some convenient temperatures, for instance the specific
heat and fourth-order cumulant pseudotransition temperatures, as
well as at a temperature close to the bulk transition temperature
($T_{c}=1.80845$). The inspection of their time-behavior verified
the accuracy of our original estimates and convinced us that the
above described practice yields a steady improvement of the
time-accumulated average DOS of all involved methods. For large
lattices, the recorded during the WL-walk ``microcanonical
estimators'', for the application of the BH method, appeared to
suffer from some ``odd fluctuations'', possibly due to a weak
violation of  the detailed balance condition, that separated their
various thermal estimates from all other estimates obtained from
the other involved methods. Although these ``odd fluctuations''
were within the error-range, we have excluded the corresponding
estimates from our thermal averages and we have also used only the
($E,M$)-histograms recorded via the Lee-walk for the further
refinement of our susceptibility estimates. The new averages were
used as refined estimates of all properties studied and are
presented in our figures using the range $L=40-240$.

Closing this Section, let us consider two different definitions
for the order-parameter of the model under study. First, with the
help of four sublattices of the SAF ordering one may define a
two-component order-parameter and finally use its root-mean-square
(\emph{rms})~\cite{binder80}
\begin{eqnarray}
\label{eq:6}
M^{(1)}&=&\left\{M_{1}+M_{2}-(M_{3}+M_{4})\right\}/4\nonumber\\
M^{(2)}&=&\left\{M_{1}+M_{4}-(M_{2}+M_{3})\right\}/4\nonumber\\
M^{(rms)}&=&\sqrt{\left(M^{(1)}\right)^{2}+\left(M^{(2)}\right)^{2}}.
\end{eqnarray}
An alternative, and numerically more convenient, definition will
also be considered from the sum of the absolute values of the four
sublattice magnetizations
\begin{equation}
\label{eq:7} M=\sum_{i=1}^{4}\left|M_{i}\right|/4.
\end{equation}
The resulting behavior is very similar and in particular the
finite-size extensions of the resulting CrMMS completely coincide.
Therefore, for large lattices only the second order-parameter was
used in order to minimize computer memory requirements. Using now
the above definition for the order-parameter we may also define
the magnetic susceptibility $\chi$ as follows:
\begin{equation}
\label{eq:8} \chi(L)=\frac{1}{N}\left(\frac{\langle
M^{2}\rangle-\langle M\rangle^{2}}{T}\right).
\end{equation}

\section{General aspects of the first-order transition}
\label{sec:3}

In general, first-order transitions~\cite{binder87} are
characterized by discontinuities of an order-parameter, like the
internal energy or the magnetization, in an idealized infinite
system. However, in a finite-system, a true phase transition can
not occur. Instead, the jump in the order-parameter is smoothed
out into a rounded transition which becomes increasingly sharp as
the finite dimensions of the system go to infinity. In this case,
the aim of FSS theory is to estimate the width and the possible
shifts of the rounded transition and to determine the associated
scaling functions. In the following subsections we summarize the
main difficulties in making a clear distinction between first and
second-order transitions and outline useful aspects of the
existing theories of first-order transitions.

\subsection{On the nature of the phase transition for the triangular SAF model}
\label{sec:3a}

As pointed out in the introduction, Rastelli et
al.~\cite{rastelli05} considered the triangular model with
nearest- and next-nearest-neighbor antiferromagnetic interactions.
Using a conventional (Metropolis) Monte Carlo approach, these
authors concluded  that the phase transition from the ordered
phase at low temperatures to the high temperature paramagnetic
phase is first-order. The identification of the nature of the
transition was based mainly on the clear presence of the energy
double-peaks reported. Additional indications were detected from
the behavior of the peaks of the Binder's fourth-order energy
cumulant and also from the scaling of the specific heat, the
susceptibility, and the corresponding pseudotransition
temperatures. From the FSS behavior of the peaks of the Binder's
fourth-order energy cumulant they estimated limiting minimum
values different from the value of 2/3 (expected for a continuous
transition) which are typically expected for first-order
transitions. The shift behavior of the pseudotransition
temperatures was also found in agreement with what is expected at
a first-order transition ($T^{\ast}\approx T_{c}+bL^{-d}$).
However, some problems were encountered mainly with the FSS of the
specific heat peaks. The estimated divergences were found in
marked difference with the expectations of the FSS theory of
first-order transitions. For the case studied here (corresponding
to the interaction ratio $R=1$) the specific heat exponent
$\alpha/\nu$ ($C^{\ast}\approx bL^{\alpha/\nu}$) was estimated to
have a value $\sim 1.6$ instead of the value $d=2$, which is the
common expectation of all existing theories (see below).

A preliminary qualitative comparative study was recently presented
by the present authors~\cite{malakis07} where the behavior of the
present triangular SAF model was contrasted to the analogous
square SAF model. In this latter study, the marked difference
between thermally driven first- and second-order transitions was
illustrated by using the rather traditional method of observing
differences in the energy's and order-parameter's
cumulant-behavior between the two models. For the triangular model
the behavior was indicative of a first-order transition, in
agreement with Ref.~\cite{rastelli05}. For the square model
($R=1$) the behavior was strongly indicating a second-order
critical point~\cite{malakis06,malakis07}, in disagreement with
the mean-field prediction of Lopez et
al.~\cite{lopez93,lopez94,lopez99}  of a first-order transition
for $R<1.2$. Consequently, the presence of the energy double-peaks
and the scaling shift behavior $L^{-d}$, in the case of the
triangular SAF model~\cite{rastelli05,malakis07}, are indications
of a first-order transition, while the absence of energy
double-peaks and the FSS with critical exponents different from
the lattice dimensionality, in the case of square SAF
model~\cite{malakis06,malakis07}, are indications of second-order
phase transition. However, caution should be paid in both cases to
ensure that the lattice size is sufficiently large so that
irrelevant fields have scaled away and the observed behavior is
the true asymptotic one and not a strong finite-size effect that
may cease to exist in the thermodynamic limit. At this point we
may recall some already known pitfalls. The existence of the
energy double-peaks for the finite lattice Baxter-Wu
model~\cite{martinos05,schreider05} is a counter-example since
this model undergoes a second-order phase transition. Such
continuous phase transitions with a convex dip in the
microcanonical entropy, including the Baxter-Wu model and the
$q=4$ Potts model in two dimensions, have been discussed recently
by Behringer and Pleimling~\cite{behringer06}. In these cases the
pseudosignatures of ``first-order transitions'' are finite-size
effects. Similarly the apparent ``first-order transition''
observed in the fixed magnetization version of the Ising model is
also a finite-size effect~\cite{pleimling01,binder03,martinos06}
and it has been shown that this ``transition'' ceases to exist in
the thermodynamic limit.

Hence, in order to provide strong evidence and to elucidate the
distinctive features for models undergoing first-order
transitions, accurate and detailed numerical studies are
necessary. Such studies have been carried out on the $q=5$, $q=8$,
and $q=10$ Potts model and most of the existing theories have been
tested and verified using this
model~\cite{lee90,lee91,challa86,borgs92a,janke93}. The
demonstration of the first-order transition features is more
difficult, as should be expected, for weak transitions (the $q=5$
Potts model is a well-known example) since in these cases the
correlation length is relatively large and strong finite-size
effects may obscure the true asymptotic behavior~\cite{lee91}. The
present extensive numerical study, is an attempt to repeat some of
these tests on the triangular SAF model, which as will be seen is
an interesting weak first-order transition. Our  accurate
numerical data will permit us to adequately determine the features
of the first-order transition of the model, to make interesting
comparisons with previous studies, and to draw conclusions related
to the applicability of the existing FSS theories.

\subsubsection{The Lee-Kosterlitz method and the latent heat of the transition}
\label{sec:3aa}

We proceed in interpreting our numerical data using the free
energy barrier method proposed by Lee and
Kosterlitz~\cite{lee90,lee91}. With this method one may
unambiguously identify the order of the transition and also
evaluate the latent heat of the transition. By computing the
energy distribution at various temperatures, we locate a
pseudotransition temperature $T_{h}$, at which the two
equal-height peaks of the energy distribution are observed. These
two peaks are located at the values of the energies per site
$e_{o}(L)=E_{o}(L)/L^{d}$ and $e_{d}(L)=E_{d}(L)/L^{d}$,
corresponding to the ordered and disordered states respectively,
and are separated by a minimum at $e_{min}(L)=E_{min}(L)/L^{d}$.
Following Lee and Kosterlitz~\cite{lee91}, the free energy barrier
is defined with the help of the microcanonical free
energy~\cite{lee91,janke93}, $F(e,L)\equiv-\ln{P(e,L)}$, where
$P(e,L)$ is the energy distribution at the pseudotransition
temperature $T_{h}$. At a first-order transition the
microcanonical free energy is assumed to have an expansion of the
form
\begin{equation}
\label{eq:9} F(e,L)=L^{d}f(e)+L^{d-1}f_{\sigma}(e)+\cdots,
\end{equation}
where $f(e)$ is the bulk free energy which is a minimum and
constant for $e_{o}\leq e \leq e_{d}$ and $f_{\sigma}(e)$ is a
surface free energy which has a maximum at $e_{min}$. The
arguments presented by Lee and Kosterlitz~\cite{lee91} suggest the
following scaling forms for the finite-size estimates of the
locations of the equal-height peaks
\begin{equation}
\label{eq:10} e_{o}(L)=e_{o}-\mathcal{O}(L^{-1})
\end{equation}
and
\begin{equation}
\label{eq:11} e_{d}(L)=e_{d}+\mathcal{O}(L^{-1}),
\end{equation}
where $e_{o}$ and $e_{d}$ are the bulk energies of the ordered and
disordered states respectively. From the above assumptions for the
bulk and surface free energies, it follows that the maximum of the
surface energy at $e_{min}$ produces a free energy barrier that
scales according to
\begin{equation}
\label{eq:12} \Delta F(L)=F(e_{min},L)-F(e_{o},L)\sim L^{d-1}.
\end{equation}
This scaling form has been used as a basic
test~\cite{lee90,lee91,janke93,ledue98} for detecting a
temperature-driven first-order transition, while the predicted
scaling behavior [Eqs.~(\ref{eq:10}) and (\ref{eq:11})] can be
used to extrapolate and obtain the latent heat of the transition
$\Delta e=e_{d}-e_{o}$.

Fig.~\ref{fig:1} presents the energy distributions for small and
large lattice sizes. Although, for the larger lattices the
simulation error fluctuations are evident, it is also very clear
that with increasing lattice size the barrier between the two
peaks is steadily increasing, signaling the emergence of the
expected two delta-peak behavior in the thermodynamic limit. Thus,
this figure is an effective demonstration of the persistence of
the first-order transition. Fig.\ref{fig:2}(a) is the
corresponding illustration, including all lattice sizes studied,
of the behavior of the free energy barrier, defined above. It is
apparent from this figure that the surface tension, that is the
quantity $\Delta F/L=\{k_{B}T\ln{(P_{max}/P_{min})}\}/L$, tends to
a non-zero value and the attempt to extrapolate by the dotted line
gives a value of the order 0.00539(12). Thus, the expectation
(\ref{eq:12}) of the arguments of by Lee and
Kosterlitz~\cite{lee91} is well verified. Fig.~\ref{fig:2}(b)
illustrates the $L$-dependence of the location of the energy
density peaks (minima of the free energy) as well as their
difference. Again in the scale shown, the expectations described
by Eqs.~(\ref{eq:10}) and (\ref{eq:11}) are well satisfied. Thus,
using linear extrapolations in the range of large $L$, we have
estimated the bulk energies $e_{o}=-1.61784(728)$ and
$e_{d}=-1.48852(565)$, and the latent heat of the transition
$\Delta e=e_{d}-e_{o}=0.129(12)$. This is to be compared with the
value $0.176(6)$ estimated by Rastelli et al.~\cite{rastelli05}.

\subsection{The double Gaussian approximation}
\label{sec:3b}

K. Binder and co-workers~\cite{binder84,challa86,binder81}
introduced the so-called double Gaussian approximation. This
assumption concerns the probability distribution of the
order-parameter and has become a widely used feature in the
theories of first-order transitions. The original approach leading
to this approximation was based on the intuitive extension of the
Gaussian approximation of a single phase, which may be
theoretically justified by appealing to the central limit theorem.
It was argued~\cite{binder84,challa86,binder81} that, the
distinctive feature of a first-order transition is the coexistence
of phases at the transition point and in fact this important
observation is the starting point of other theories of first-order
transitions~\cite{lee90,lee91,fisher82,binder84,challa86,borgs92a,janke93,binder81,blote81,privman83,borgs90,borgs91,borgs92b,borgs92c,borgs92d,pirogov75,zahradnik84,borgs89}.
Ignoring effectively interface effects between domains of
different phases, the equilibrium of the phases is described by a
superposition of Gaussians centered at different values of the
order parameter corresponding to the different coexisting phases.

Furthermore, it was soon realized~\cite{lee90,lee91} and
theoretically~\cite{borgs90,borgs91,borgs92b,borgs92c,borgs92d,pirogov75,zahradnik84,borgs89}
justified, that the same qualitative picture can be obtained by a
more fundamental assumption. The partition function of a large
finite system, with periodic boundary conditions, undergoing in
the limit $L\rightarrow \infty$ a first-order transition, is well
approximated at the transition point as a sum of terms, each
describing a separate phase. For temperature-driven first-order
phase transitions, in which $q$ equivalent ordered phases coexist
at the transition temperature ($\beta_{c}$) with one disordered
phase, the dominant contributions in the partition function,
separated in contributions from the different coexisting phases,
may be written as~\cite{lee91}
\begin{equation}
\label{eq:13}
Z=\sum_{i=1}^{q+1}e^{-\beta_{c}f_{i}(\beta_{c})L^{d}},
\end{equation}
where $f_{i}(\beta_{c})$ is the free energy per site of the
\emph{ith} phase in the thermodynamic limit. This fundamental
relationship is a statement that each phase contributes equally to
$Z$ at the transition temperature
($f_{i}(\beta_{c})=f(\beta_{c})$). As pointed out by Lee and
Kosterlitz~\cite{lee91}, the double Gaussian approximation may be
obtained by assuming an extension of (\ref{eq:13}) in the vicinity
of $T_{c}$ and using appropriate temperature Taylor expansions for
the free energies $f_{i}(\beta)$ of the coexisting phases. The
probability density for the energy can then be obtained by the
inverse Laplace transform of the expanded partition function and,
in order $\mathcal{O}(L^{-d})$ in the free energy expansion, one
can derive the double Gaussian form
\begin{equation}
\label{eq:14}
P(e)=\frac{e^{\Delta}}{e^{\Delta}+e^{-\Delta}}\Phi_{G}(e;\widetilde{e}_{o},\sigma_{o})+
\frac{e^{-\Delta}}{e^{\Delta}+e^{-\Delta}}\Phi_{G}(e;\widetilde{e}_{d},\sigma_{d}),
\end{equation}
where $\Phi_{G}(e;\widetilde{e}_{i},\sigma_{i})$ is the normal
distribution with mean
$\widetilde{e}_{i}=e_{i}+C_{i}(\frac{T_{c}}{T})(T-T_{c})$ and
variance $\sigma_{i}^{2}=k_{B}T_{c}^{2}C_{i}L^{-d}$. $e_{i}$ and
$C_{i}$ are the energy and specific heat of the \emph{ith} bulk
phase, derived as appropriate derivatives from the corresponding
free energies. Furthermore,
\begin{equation}
\label{eq:15}
\Delta=\frac{1}{2}\ln{q}+\frac{1}{2}(e_{o}-e_{d})\frac{T_{c}}{T}(T-T_{c})L^{d}+
\frac{1}{4}(C_{o}-C_{d})\left(\frac{T-T_{c}}{T}\right)^{2}L^{d},
\end{equation}
so that the ratio of weights corresponding to the two terms of
Eq.~(\ref{eq:14}) satisfies the relation
\begin{equation}
\label{eq:16}
\frac{w_{o}}{w_{d}}=e^{2\Delta}=qe^{\beta(f_{d}-f_{o})L^{d}}.
\end{equation}

From the above distribution, or in a more straightforward way, by
using differentiation of the expanded partition function one can
deduce the finite-size behavior of all energy cumulants, as shown
by Lee and Kosterlitz~\cite{lee91}. These tedious calculations
have been also repeated by Janke~\cite{janke93} and in the
following we summarize only the relevant results that have been
used in our study of the triangular SAF model. At this point let
us emphasize the fact, stated explicitly bellow, that the double
Gaussian approximations predicts, in general, that all finite-size
contributions enter in the scaling equations in powers of
$L^{-d}$. The general shift behavior predicted for the various
pseudotransition temperatures (denoted collectively as
$T_{i}^{\ast}$) is
\begin{equation}
\label{eq:17} T_{i}^{\ast}=T_{c}+bL^{-d}(1+\mathcal{O}(L^{-d})),
\end{equation}
where these temperatures correspond to the peaks of various energy
cumulants such as those defined bellow. As noted above, not only
the main shift contribution but also the higher-order corrections
enter in Eq.~(\ref{eq:17}) only in powers of $L^{d}$. Three
different energy cumulants have been used in our calculations, the
specific heat per site $C(L)$, the second-order cumulant
$U_{2}(L)$, and the Binder's fourth-order cumulant $V_{4}(L)$.
Bellow we give their definitions and the predicted scaling
behavior~\cite{lee91,janke93} at the corresponding
pseudotransition temperature, at which the cumulant peak (maximum
or minimum) occurs
\begin{equation}
\label{eq:18} C(L)=L^{-d}\left\{\langle E^{2}\rangle-\langle
E\rangle^{2}\right\}/T^{2}
\end{equation}

\begin{equation}
\label{eq:19}
C(L)L^{-d}|_{max}=\frac{(e_{o}-e_{d})^{2}}{4T_{c}^{2}}(1+\mathcal{O}(L^{-d}))
\end{equation}

\begin{equation}
\label{eq:20} U_{2}(L)=\frac{\langle E^{2}\rangle}{\langle
E\rangle^{2}}
\end{equation}

\begin{equation}
\label{eq:21}
U_{2}(L)|_{max}=\frac{(e_{o}+e_{d})^{2}}{4e_{o}e_{d}}+\mathcal{O}(L^{-d})
\end{equation}

\begin{equation}
\label{eq:22} V_{4}(L)=1-\frac{\langle E^{4}\rangle}{3\langle
E^{2}\rangle^{2}}
\end{equation}

\begin{equation}
\label{eq:23}
V_{4}(L)|_{min}=\frac{2}{3}-\frac{(e_{o}/e_{d}-e_{d}/e_{o})^{2}}{12}+\mathcal{O}(L^{-d}).
\end{equation}

In the following Section we proceed to test the above scaling laws
for the triangular SAF model. In particular, the coefficient of
the dominant diverging term of the specific heat and the limiting
values of the cumulants $U_{2}$ and $V_{4}$, will be determined in
two independent ways: (a) by finding their values from convincing
fitting assumptions of our peak-data and (b) from the extrapolated
values of the energies $e_{o}$ and $e_{d}$ (see
Fig.~\ref{fig:2}(b) and Eqs.~(\ref{eq:10}) and (\ref{eq:11})) and
the above predictions. The proposed comparison will provide a
useful self-consistency test of our results but also of the
theoretical predictions. It will also assist a critical discussion
on the higher-order corrections of the double Gaussian
approximation. Noteworthy, that the above leading terms may also
be derived heuristically from a simple two-phase model, as shown
by Janke~\cite{janke93}.

\subsection{Number-of-phases parameters and the determination of transition points}
\label{sec:3c}

Traditionally, the general shift behavior $T^{\ast}\approx
T_{c}+bL^{-d}$ has been used for both the identification of a
first-order transition and the determination of the transition
temperature. However, Borgs and Janke~\cite{borgs92a} have
suggested additional methods facilitating the determination of
transition points for systems with periodic boundary conditions.
Two such methods will be discussed below and then elaborated in
the sequel to observe the behavior of the present first-order
transition. In the first method, a parameter is introduced that
requires two different finite lattices of volumes
$V_{1}=L_{1}^{d}$ and $V_{2}=L_{2}^{d}$ respectively. This
parameter has been called the number-of-phases
parameter~\cite{janke93} and is defined as:
\begin{equation}
\label{eq:24}
N(V_{1},V_{2};\beta)=\left[\frac{Z_{per}(V_{1};\beta)^{\alpha}}{Z_{per}(V_{2};\beta)}\right]^{1/(\alpha-1)},
\end{equation}
where $\alpha$ is the volume ratio
$\alpha=V_{2}/V_{1}=(L_{2}/L_{1})^{d}$. The origin of this
parameter is the fundamental relationship (\ref{eq:13}), which is
extended, as earlier, in the vicinity of the transition point by
introducing~\cite{borgs92a,janke93} metastable free energy
densities $f_{i}(\beta)$. These are defined in such a way that are
equal to the bulk free energy $f(\beta)$ in the temperature range
in which the corresponding phase is a stable phase and strictly
larger than $f(\beta)$ when the phase is unstable. The coexistence
partition function is now written as~\cite{borgs92a,janke93}
\begin{equation}
\label{eq:25} Z_{per}(V;\beta)=\left\{\sum_{i=1}^{q+1}e^{-\beta
f_{i}(\beta)V}\right\}\left(1+\mathcal{O}(Ve^{-L/L_{o}})\right).
\end{equation}
A heuristic derivation of the above exponential correction bound
has been presented in Ref.~\cite{borgs92a}. The hypothesis for
exponential small shifts, to be discussed in the next Section,
derives from the above bound. Eq.~(\ref{eq:25}) and the
assumptions of the behavior of metastable free energy densities
may be used to show~\cite{janke93} that the parameter
\begin{equation}
\label{eq:26} N(\beta)=\lim_{V\rightarrow \infty}
Z_{per}(V;\beta)e^{\beta f(\beta)V}
\end{equation}
is equal to the number of stable phases, which with increasing
temperature takes the values $q$, $q+1$, and $1$. A finite-size
approximation of the above parameter will be expected to have a
peak at a finite-size transition point ($T_{V/V}$) and the
number-of-phases parameter defined in Eq.~(\ref{eq:24}) is a
suitable such approximation that may be used to locate the
transition point. The condition for the maximum of the parameter
of Eq.~(\ref{eq:24}) corresponds to the crossing point of the mean
energy functions of the pair of lattices involved and reads as:
\begin{equation}
\label{eq:27} \alpha \langle E_{1}\rangle|_{T_{V/V}}=\langle
E_{2}\rangle|_{T_{V/V}}\;\;\; or \;\;\;\langle
e_{1}\rangle|_{T_{V/V}}=\langle e_{2}\rangle|_{T_{V/V}}.
\end{equation}
The location of the peaks of the parameter (\ref{eq:24}) can be
estimated by a dominant CrMES scheme, such as that followed by the
present authors in other
applications~\cite{malakis06,malakis04,malakis05}. Of course, the
restricted energy space scheme should be sufficient for both
lattices of the pair in the temperature range of interest, that is
in the region of the peaks. We shall use the sequence of pairs of
lattices: ($40$, $80$), ($50$, $100$), ($60$, $120$), ($70$,
$140$), ($80$, $160$), ($100$, $200$), and ($120$, $240$), with
volume ratio $\alpha=4$ ($L_{2}=2L_{1}$, $d=2$).
Fig.~\ref{fig:3}(a) illustrates the behavior which is similar with
the behavior presented by Janke~\cite{janke93} for the Potts
model.

We now discuss a variant of the above number-of-phases parameter.
First, we define a ``reduced partition function''. The maximum
term, at each temperature, has been factorized and dropped out
from the partition function's sum. Thus,
\begin{equation}
\label{eq:28}\widetilde{Z}(V;\beta)=\sum_{(\widetilde{E}_{-},\widetilde{E}_{+})}e^{\widetilde{\Phi}(E)},
\end{equation}
where $\widetilde{E}=\widetilde{E}(T)$ is the most probable energy
corresponding to the maximum term, and
\begin{equation}
\label{eq:29}\widetilde{\Phi}(E)=S(E)-\beta
E-[S(\widetilde{E})-\beta \widetilde{E}].
\end{equation}
Since the most probable energy has two values at the equal-height
temperature ($T_{h}$), the above reduction introduces in the above
definition microcanonical features of the transition. These
features are now carried over to the new ratio parameter
\begin{equation}
\label{eq:30}\widetilde{N}_{\lambda}(V_{1},V_{2};\beta)=\frac{\widetilde{Z}(V_{1};\beta)^{\lambda}}{\widetilde{Z}(V_{2};\beta)}.
\end{equation}
Let us regard the exponent $\lambda$ as a free parameter, not
necessarily equal to the volume ratio of the two systems
($\alpha=4$), and observe the resulting behavior.
Fig.~\ref{fig:3}(b) shows the behavior of this new ratio parameter
for the case $\lambda=2$ and the three larger lattice size pairs
of the sequence considered above. It can be seen from this figure
that by increasing the temperature this parameter shows a first
maximum peak ($T_{max1}$) which resembles the peak of the original
number-of-phases parameter. Further increase in the temperature
yields, for each pair, two additional very sharp peaks, one
minimum ($T_{min}$) and one maximum ($T_{max2}$). These two peaks
are reflections of the maximum term reduction in the definition
(\ref{eq:30}). The sharp minimum corresponds to the equal-height
pseudotransition temperature $T_{h}$ of the larger lattice, while
the sharp maximum corresponds to the equal-height pseudotransition
temperature $T_{h}$ of the smaller lattice. The value of the
exponent $\lambda$ determines the location of the first maximum
peak and the sharpness of these graphs. Thus, a similar but less
pronounced picture is obtained for the values $\lambda=3$ and
$\lambda=4$, and it is found that the location of the two new
sharp peaks are not notably influenced by the change in the value
of the exponent $\lambda$. This can be easily understood since the
new sharp peaks are in effect properties of the two separated
lattices. The peaks of this new ratio can be also be described by
an approximate crossing condition, similar to Eq.~(\ref{eq:27}),
which may be written as:
\begin{equation}
\label{eq:31}\lambda(\langle
e_{1}\rangle-\widetilde{e}_{1})|_{\widetilde{T}}=\alpha(\langle
e_{1}\rangle-\widetilde{e}_{1})|_{\widetilde{T}}.
\end{equation}
In the above condition we have neglected the temperature variation
of the most probable energy value $\widetilde{E}$. This variation
will produce vanishing contributions, as tested also numerically,
involving the difference between microcanonical and canonical
temperature $[(\vartheta S_{i}/\vartheta
E_{i})|_{\widetilde{E}_{i}}-\beta]$.

Fig.~\ref{fig:4}(a) is an illustration of the crossing point of
the mean energy functions of a pair of systems ($L_{1}=80$,
$L_{2}=160$) corresponding to the peak of the original
number-of-phases parameter (\ref{eq:24}). In this the behavior of
the most probable energies of the pair of systems is also
illustrated. Fig.~\ref{fig:4}(b) presents the temperature behavior
of the differences, $\langle e\rangle-\widetilde{e}$, of the two
characteristic energies for each system of the chosen pair. For
the larger lattice the graph corresponding to the double of the
difference $\langle e\rangle-\widetilde{e}$ is also shown in order
to facilitate the illustration of the above crossing relationship
(\ref{eq:31}) for $\lambda=2$. Noteworthy, that this illustration
provides an alternative practical way of locating the equal-height
pseudotransition temperature $T_{h}$, since at this temperature
(inversion point) the corresponding difference changes sign. As we
increase the temperature, the difference curve of the smaller
lattice is crossed by the curve of the larger lattice at a point
($T_{max1}$) corresponding to the first maximum in
Fig.~\ref{fig:3}(b). As can be seen the location of this point
($\widetilde{T}_{\lambda}$) depends on the value of the exponent
$\lambda$ and for the value $\lambda=\alpha/2=2$, the crossing is
deeper and occurs at a lower temperature, than in the case
corresponding to $\lambda=\alpha=4$. This explains why the first
maximum peak ($T_{max1}$) in the case $\lambda=2$ is more
pronounced as mentioned earlier. Finally, the sharp minimum
($T_{min}$) and the sharp maximum ($T_{max2}$) in
Fig.~\ref{fig:3}(b) are described by the crossing points
corresponding to the inversion points of the larger and smaller
lattice respectively and their location is not practically
influenced by the value of the exponent $\lambda$.

We shall conclude this Section with a brief reference on the
second definition of a finite-size transition point introduced by
Borgs and Janke~\cite{borgs92a}. This equal-weight finite-size
transition point ($T_{W}$) requires data from one lattice only and
is the point where the ratio of the total weight of the ordered
phases to the weight of the disordered phase approaches $q$, so
that
\begin{equation}
\label{eq:32}R(V,E)=\sum_{E<E_{m}}P(E)/\sum_{E\ge E_{m}}P(E)\equiv
\frac{W_{o}}{W_{d}}|_{T_{W}}=q,
\end{equation}
where $E_{m}$ may be chosen in various ways and following
Refs.~\cite{borgs92a,janke93} we will choose it to be the energy
of the minimum between the two equal-height peaks at the
pseudotransition temperature $T_{h}$. As a variant of the above
definition, we have attempted to satisfy Eq.~(\ref{eq:32}) using
the part of the weights above the minimum between two unequal
peaks. The resulting transition points have similar behavior and
will not be presented.

\section{Analysis of numerical data}
\label{sec:4}

\subsection{Estimation of the transition temperature}
\label{sec:4a}

Using our accurate data and various alternative methods we have
estimated the transition temperature with an error at most in the
fifth significant figure. Our safe estimate is $T_{c}=1.8085(1)$
and is an improvement of the three estimates of the order of
$T_{c}=1.8078(1)$, given originally by Rastelli et
al.~\cite{rastelli05} and close to their magnetic fourth-order
cumulant estimate $T_{c}=1.8084(1)$. An interface method
estimation of this transition point ($T_{c}=2.044$) was given by
Slotte and Hemmer~\cite{slotte84}. Fig.~\ref{fig:5}(a) presents a
first attempt to estimate the transition temperature by a
simultaneous fitting on five pseudotransition temperatures. As
indicated, in an obvious notation on the graph, these are the
equal-height temperature ($T_{h}$) and the pseudotransition
temperatures corresponding to the peaks of the specific heat,
second- and fourth-order energy cumulants, and finally the peak of
the susceptibility. In the first attempt the shift behavior
predicted by the double Gaussian approximation has been assumed
including also the first shift correction term $L^{-2}$ (i.e.
$\mathcal{O}(L^{-d})$). Fig.~\ref{fig:5}(b) presents a second
attempt using now a variant of the above assumption by replacing
the shift correction term with an $L^{-1}$ correction. This
replacement makes practically no difference in estimating the
transition temperature because the dominant shift term is about
three orders of magnitudes larger than the next correction in all
cases, and the asymptotic behavior has been well reached in the
temperature shift behavior. Noteworthy that, if we restrict the
fitting attempt in the range $L=30-140$, we will obtain an
estimate $T_{c}=1.80845(5)$.

Next we present the behavior of the new transition point
observables introduced by Borgs and Janke~\cite{borgs92a}, as well
as the similar ones introduced and discussed in the last
subsection. Although we will not try to further refine the above
accurate estimate, it is of interest to illustrate their behavior,
show that they lead to the same estimate, and discuss their
behavior in comparison with previous studies on the Potts model.
Fig.~\ref{fig:6}(a) illustrates the behavior of the three peaks of
the reduced number-of-phases parameter defined in the last Section
in Eq.~(\ref{eq:30}) for $\lambda=2$. The data for the two peaks
($T_{max2}$ and $T_{min}$) are approaching from above the
transition point and coincide in practice with the equal-height
temperatures for the smaller and larger lattices respectively, as
explained in subsection~\ref{sec:3c}. The data for the peaks
$T_{max1}$, which resemble the original peaks of the
number-of-phases parameter of Borgs and Janke~\cite{borgs92a}, are
also moving steadily to the expected limit and their fitting,
using Eq.~(\ref{eq:17}), yields again the estimate
$T_{c}=1.8085(1)$. Fig.~\ref{fig:6}(b) collects the behavior of
six finite-size transition points and compares them with our
estimate for the transition point and also with the estimate
$T_{c}=1.8078$ of Rastelli et al.~\cite{rastelli05}. We have
included in this figure two transition points depending only on
one lattice, namely the equal-height transition point ($T_{h}$)
and the weight transition point ($T_{W}$) defined in
Eq.~(\ref{eq:32}). The rest of the shown transition points are the
original number-of-phases parameter transition point introduced by
Borgs and Janke~\cite{borgs92a}, denoted as $T_{V/V}$, and the
left maximum point ($T_{max1}$ or $\widetilde{T}_{\lambda}$) of
the reduced number-of-phases parameter introduced in this paper
for three values of the exponent $\lambda$, $\lambda=2,3$ and $4$
(see the discussion below Eq.~(\ref{eq:31})). Their limiting
behavior is an extra verification of our estimate for the
transition temperature.

Borgs and Janke~\cite{borgs92a} in their Fig. 2, and
Janke~\cite{janke93} in his Fig. 11, have illustrated the behavior
of the transition points $T_{V/V}$ and the $T_{W}$ for the Potts
model with $q=5$, $q=8$, and $q=10$. As it can be seen from the
mentioned figures, the $q=8$ and $q=10$ finite-size transition
points approach their bulk limit for relatively small lattices.
Therefore, these authors have concluded that for strong
first-order phase transitions the exponential bound in
Eq.~(\ref{eq:25}) provides a natural explanation of the observed
``exponentially small'' shifts with respect to the infinite-volume
transition point. For the weak first-order transition of the $q=5$
Potts model, the weight transition point $T_{W}$ can also be seen
from the mentioned figures to have reached the bulk limit even for
the smallest lattice size $L=20$. However, the number-of-phases
parameter transition points $T_{V/V}$ did not obey this
exceptional behavior, but rather its shift behavior resembled the
shift behavior of the traditional pseudotransition temperatures,
i.e. an $L^{-d}$ shift. Accordingly, Janke~\cite{janke93} has
raised also the question of a possible fortuitous cancellation and
has suggested the interest to the investigation of these
exponentially small shifts in other weak first-order transitions.
Fig.~\ref{fig:6}(b) provides an example showing that for the
present weak first-order transition the expectation for
exponentially small shifts is not verified for both transition
points $T_{W}$ and $T_{V/V}$. Their behavior differs appreciably
from their bulk limit and the reduced number-of-phases parameter
for the value $\lambda=4$ yields the smaller shifts, smaller than
that of the natural number-of-phases parameter.

\subsection{Behavior of the energy cumulants and the magnetic susceptibility}
\label{sec:4b}

It is well known that the FSS behavior of the specific heat, and
in particular its peak-scaling behavior, may be obscured by the
presence of unknown correction terms and the extraction of the
scaling exponent may be a numerically indecisive and difficult
task~\cite{newman99,landau00,ledue98}. The accuracy of numerical
data is then of crucial importance~\cite{landau00}. This
unpleasant problem may also occur in the estimation of other
related energy cumulants. For the present model such problems were
obviously encountered by Rastelli et al.~\cite{rastelli05}. Their
estimation of the specific heat exponent $\alpha/\nu$
($C^{\ast}\approx bL^{\alpha/\nu}$) was found in clear
disagreement (estimated to have a value $\sim 1.6$) with the
theoretically expected value $\alpha/\nu=d=2$. Since the evidence
for the present first-order transition is now overwhelming we will
adopt the generally accepted theoretical prediction for the
leading contribution to the specific heat divergence. Thus, we
shall fix the value of the exponent $\alpha/\nu=d=2$ and we will
examine whether there is a stable form of corrections terms that
is compatible with our accurate data in the peak-region.

Fitting our numerical data from the first $10$ WL runs in various
$L$-ranges, we found that the simple correction term of the form
$cL^{-1}$, instead of $cL^{-2}$ predicted by the double Gaussian
approximation, yielded stable fittings in all the $L$-ranges
tried, from $L=30$ to $L=480$. The very small variations in the
involved coefficients ($5\%$ or less) of the fitting attempts are,
of course, strong indications in favor of this otherwise at hoc
assumption. Fig.~\ref{fig:7} shows that even the remote
$L$-ranges: $L=30-140$ and $L=100-480$ produce $4\%$ deviation in
the coefficient of the leading term and only $1\%$ in the
coefficient of the assumed correction term. Fig.~\ref{fig:8}
provides further strong support in favor of the above assumption
using our second-stage refined data in the range $L=40-240$. This
figure present simultaneous fittings of the specific heat data not
only at its pseudotransition temperature but also in the other
pseudotransition temperatures corresponding to the peaks of the
other energy cumulants, the peak of the magnetic susceptibility
and the equal-height transition point. Noteworthy, that such a
simultaneous fitting approach, quite easily implemented within our
entropic scheme, would be a rather demanding task in a traditional
importance sampling Monte Carlo scheme. In this way we are probing
systematically the FSS behavior of the divergence of the specific
heat in a wide region around its peak. Fig.~\ref{fig:8}(a)
illustrates the fitting attempt based on the above mentioned
($cL^{-1}$) assumption and Fig.~\ref{fig:8}(b) shows a remarkable
decline from the double Gaussian prediction ($cL^{-2}$). The
double Gaussian prediction does not correctly describe the
correction term and this is reflected in the resulting very large
$\chi^{2}$ error value which is about $35$ times larger than the
value corresponding to the well obeyed correction term
($cL^{-1}$). Let us now use the corresponding coefficients of the
dominant divergences for the peaks of the specific heat to compute
the resulting values for the latent heat and compare these values
with the value obtained earlier from the extrapolations of
Fig.~\ref{fig:2}(b), based on the proposal of Lee and
Kosterlitz~\cite{lee91} and described in Eqs.~(\ref{eq:10}) and
(\ref{eq:11}). The resulting values for the latent heat from the
fittings in Fig.~\ref{fig:8} are in the range $0.121-0.133$ and
$0.143-0.163$ respectively. These values should be compared with
the value 0.129(12), see also Table~\ref{tab:1} below, obtained
from the extrapolated values of $e_{o}$ and $e_{d}$.

An analogous behavior is observed in the FSS cbehavior of the
other two energy cumulants. Fig.~\ref{fig:9}(a) shows in the same
graph fitting attempts using the data for the minima of the
Binder's fourth-order cumulant presuming again the two different
assumptions for the correction terms ($bL^{-2}$ and $bL^{-1}$). As
can be seen from this figure the double Gaussian prediction fails
again to correctly describe the correction term. The resulting
very large $\chi^{2}$ error value is now $135$ times larger than
the value corresponding to the simple correction term of the form
$bL^{-1}$. Furthermore, the limiting value corresponding to the
assumption using the double Gaussian correction term is
$0.66259(40)$ for the $L$-range shown, which differs appreciably
from the value $0.66435(33)$, obtained from Eq.~(\ref{eq:23}) and
the extrapolated values of $e_{o}$ and $e_{d}$ given in
Table~\ref{tab:1}. On the other hand the limiting value resulting
from a leading correction of the form $L^{-1}$ gives $0.66465(25)$
for the $L$-range shown, as illustrated in Fig.~\ref{fig:9}(b),
and favors an expansion starting with the term $bL^{-1}$.
Fig.~\ref{fig:10} repeats the above comparison for the
second-order cumulant $U_{2}$. The expansion starting with the
term $bL^{-1}$ is surprisingly well obeyed, while the double
Gaussian behavior is inconsistent. The limiting value obtained
from double Gaussian correction term is $1.00308(25)$ in clear
disagreement with the value $1.00174(25)$ obtained from
Eq.~(\ref{eq:21}) and the extrapolated values of $e_{o}$ and
$e_{d}$, while the proposed expansion in Fig.~\ref{fig:10}(b)
gives the value $1.00163(15)$ quite in agreement with the also
approximate estimate $1.00174(25)$.

Table~\ref{tab:1} summarizes the estimates for the above thermal
characteristics including also the corresponding exact
values~\cite{lee91,janke93,baxter73,kihara54} for the cases
$q=10,8,6$ and $q=5$ of the Potts model. The estimates for the
triangular SAF model with $R=1$ interpolate between the values of
the $q=5$ and $q=6$ Potts model. Therefore, we have called the
present transition a weak first-order transition. It appears that
the six-fold degeneracy of the ground state of the present model
plays an important role in determining the weakness of the
transition, and possibly the occurrence of its first-order
character. Finally, let us turn to the behavior of the magnetic
susceptibility $\chi$. We present in Fig.~\ref{fig:11} the
divergence of the susceptibility $\chi$ at the five pseudocritical
temperatures $T_{i}^{\ast}$, using again the two types of
corrections, i.e. an $L^{-1}$ (Fig.~\ref{fig:11}(a)) and an
$L^{-2}$ (Fig.~\ref{fig:11}(b)), respectively. The solid lines are
simultaneous fitting attempts, using the power laws shown in the
figures. Comparing Figs.~\ref{fig:11}(a) and (b) we see that,
although one can not just by inspection differentiate between an
$L^{-1}$ and an $L^{-2}$ correction term, $\chi^{2}$ is smaller
for the former correction.

\section{Conclusions}
\label{sec:5}

The triangular Ising model with nearest- and next-nearest-neighbor
antiferromagnetic interactions has been studied and its
first-order transition features have been clarified, when the
interaction ratio is $R=1$. We have outlined the most important
aspects of the existing theories of first-order transitions and we
have tested on this model some basic hypothesis from these
theories, comparing our results and findings with previous work on
the Potts model. Our numerical data have been used to obtain
accurate estimates for all the thermal characteristics of the
present weak first-order transition. All the generally accepted
predictions of the finite-size scaling theory for first-order
transitions, concerning transition point shifts, thermal, and
magnetic anomalies, have been well verified for the present model.

However, two of our findings for this model are not compatible
with some theoretical or phenomenological expectations. The first
of these concerns the behavior of the new transition point
observables introduced by Borgs and Janke~\cite{borgs92a} and also
some similar transition points introduced in this paper. These
finite-size transition points are suitable finite-size
approximations of the fundamental number-of-phases parameter and
it has been theoretically argued~\cite{lee91,janke93} that should
be expected to obey exponentially small shifts. This expectation
is not verified for the present weak first-order transition.

\begin{figure}[htbp]
\includegraphics{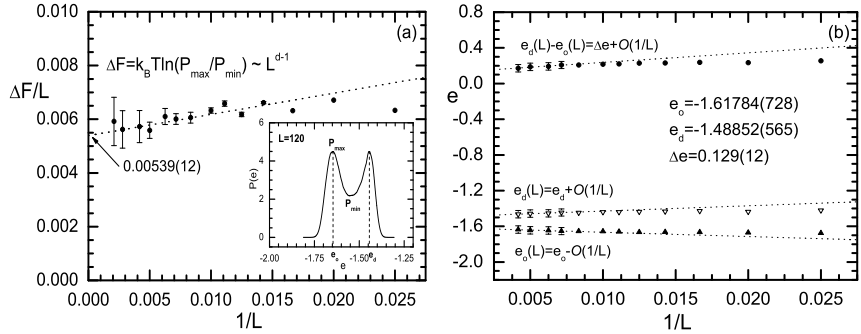}
\caption{\label{fig:2}(a) Plot of the barrier height $\Delta
F(L)L^{-d+1}$. The dotted line is an extrapolation to
$L\rightarrow \infty$, giving a nonzero value for the surface
tension of the order of $0.00539(12)$. The inset is used as a
guide for the reader. (b) The energy minima $e_{o}(L)$ and
$e_{d}(L)$ and their difference as a function of the inverse
linear size. The dotted lines are linear fits indicating the
values of the bulk energies $e_{o}$ and $e_{d}$, and that of the
latent heat $\Delta e$ at $L=\infty$.}
\end{figure}

\begin{figure}[htbp]
\includegraphics{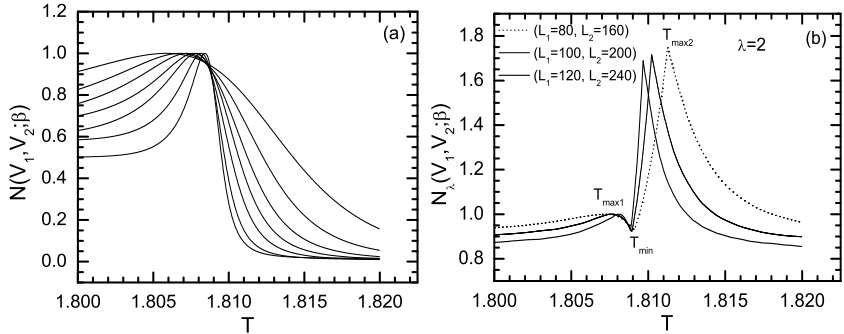}
\caption{\label{fig:3}Temperature-dependence of the ratios
$N(V_{1},V_{2};\beta)$ (a) and
$\widetilde{N}_{\lambda=2}(V_{1},V_{2};\beta)$ (b)
(Eqs.~(\ref{eq:24}) and (\ref{eq:30}), respectively). In panel (a)
all pairs are shown, i.e. ($L_{1}$, $L_{2}$)=($40$, $80$), ($50$,
$100$), ($60$, $120$), ($70$, $140$), ($80$, $160$), ($100$,
$200$), and ($120$, $240$), while in panel (b) only the pairs
($L_{1}$, $L_{2}$)=($80$, $160$), ($100$, $200$), and ($120$,
$240$) are shown.}
\end{figure}

\begin{figure}[htbp]
\includegraphics{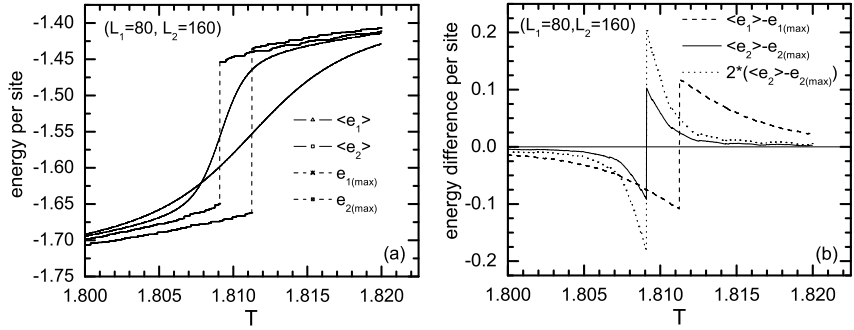}
\caption{\label{fig:4}(a) Illustration of the crossing point
corresponding to the peak of the number-of-phases parameter. The
behavior of the most probable energies of the pair of systems is
also illustrated. (b) Behavior of the differences, $\langle
e\rangle-\widetilde{e}$, of the two characteristic energies for
each system of the chosen pair. For the larger lattice the graph
corresponding to the double of the difference $\langle
e\rangle-\widetilde{e}$ is also shown in order to facilitate the
illustration of the crossing relationship (\ref{eq:31}) for
$\lambda=2$ ($\alpha=4$).}
\end{figure}

\begin{figure}[htbp]
\includegraphics{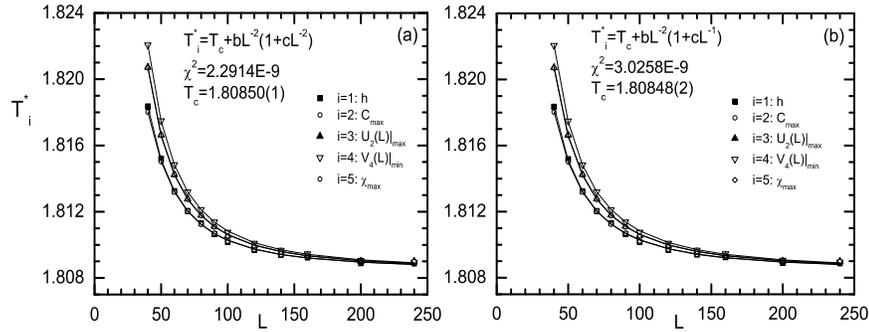}
\caption{\label{fig:5}(a) Simultaneous fittings for the shifting
of all pseudocritical temperatures $T_{i}^{\ast}$ using an
$L^{-2}$ correction. (b) The same with panel (a) but with an
$L^{-1}$ correction.}
\end{figure}

\begin{figure}[htbp]
\includegraphics{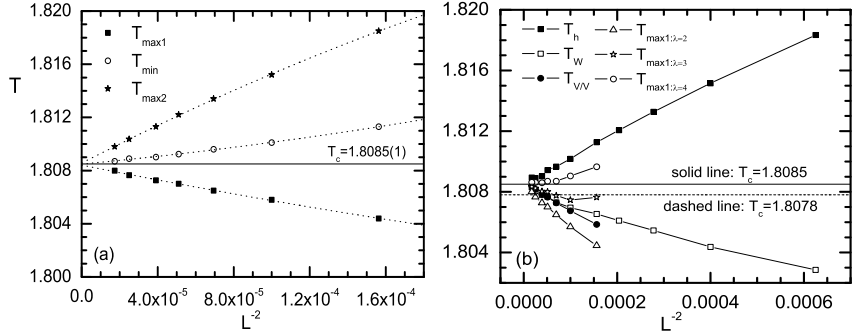}
\caption{\label{fig:6}(a) Illustration of the behavior of the
three peaks of the reduced number-of-phases parameter defined in
in Eq.~(\ref{eq:30}) for $\lambda=2$. (b) Behavior of six
finite-size transition points and comparison with our estimate
(solid line) for the transition point and also with the estimate
$T_{c}=1.8078$ of Ref.~\cite{rastelli05} (dashed line).}
\end{figure}

\begin{figure}[htbp]
\includegraphics{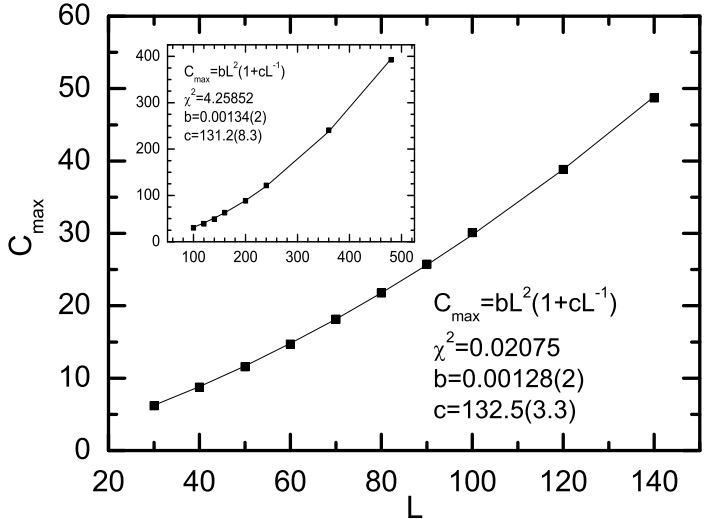}
\caption{\label{fig:7}FSS behavior of the specific heat peaks in
two lattice ranges: $L=30-140$ and $L=100-480$ (inset). An
$L^{-1}$ correction has been added to the $L^{2}$ divergence.}
\end{figure}

\begin{figure}[htbp]
\includegraphics{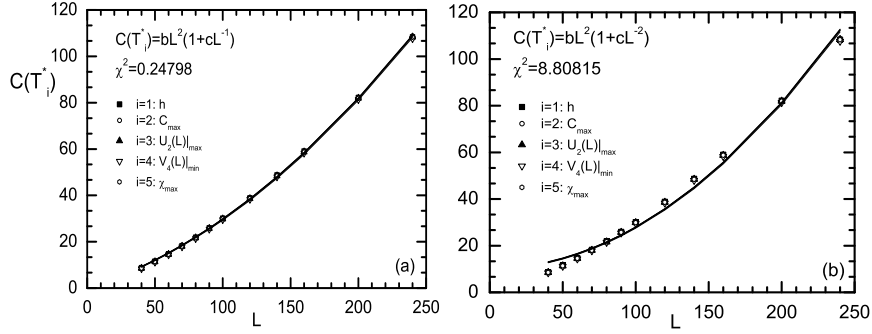}
\caption{\label{fig:8}(a) FSS behavior of the specific heat data
at the five pseudocritical temperatures $T_{i}^{\ast}$ using an
$L^{-1}$ correction. (b) The same with panel (a) but with an
$L^{-2}$ correction. Note the large difference in the $\chi^{2}$
values.}
\end{figure}

\begin{figure}[htbp]
\includegraphics{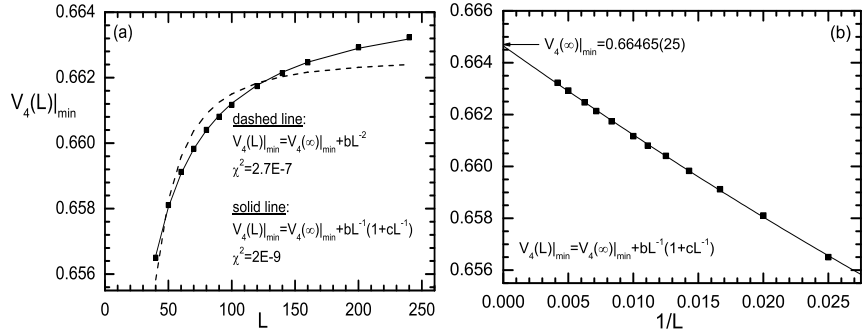}
\caption{\label{fig:9}(a) FSS behavior of $V_{4}(L)|_{min}$. Two
fitting attempts are applied as shown in the figure. $\chi^{2}$ is
much smaller for the case of an $L^{-1}$ correction. (b) FSS of
$V_{4}(L)|_{min}$, now against $L^{-1}$. The line corresponds to
the fitting formulae shown in panel (b) and gives an estimate
$V_{4}(\infty)|_{min}=0.66465(25)$.}
\end{figure}

\begin{figure}[htbp]
\includegraphics{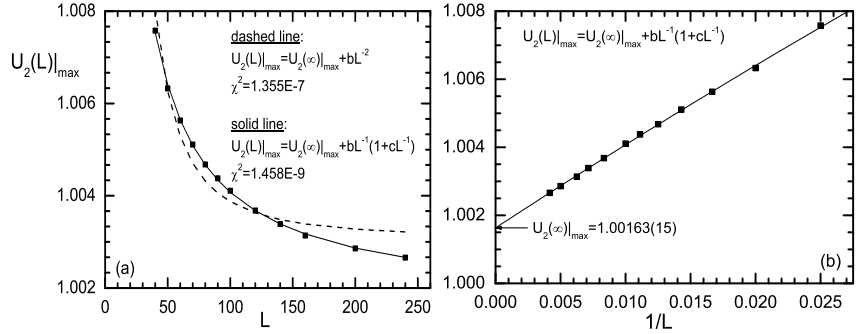}
\caption{\label{fig:10}(a) The same as in Fig.~\ref{fig:9} for
$U_{2}(L)|_{max}$. (b) FSS of $U_{2}(L)|_{max}$, against $L^{-1}$.
The solid line corresponds to the fitting formulae shown in panel
(b) and produces the estimate $U_{2}(\infty)|_{max}=1.00163(15)$.}
\end{figure}

\begin{figure}[htbp]
\includegraphics{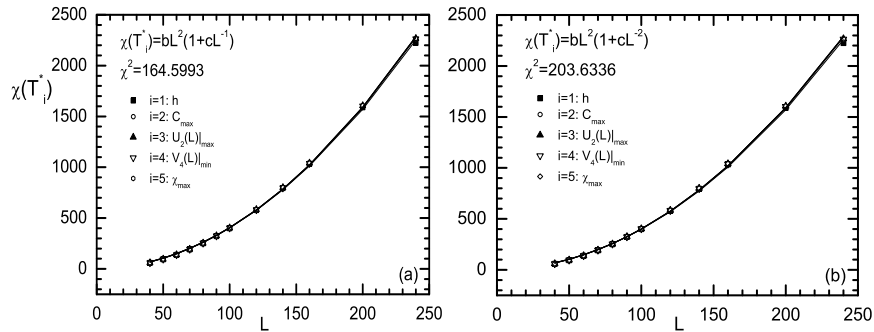}
\caption{\label{fig:11}(a) FSS behavior of the susceptibility data
at the five pseudocritical temperatures $T_{i}^{\ast}$. The solid
lines are simultaneous fitting attempts, using an $L^{-1}$
correction. (b) The same with panel (a) but with a correction of
the order of $L^{-2}$.}
\end{figure}

Finally, we have shown that the well-known double Gaussian
approximation does not describe correctly the higher correction
terms for all energy cumulants of the present model. It appears
that the first correction term in the expansions of energy
cumulants is of the form $L^{-d+1}(=L^{-1})$ and not of the form
$L^{-d}(=L^{-2})$, expected from the double Gaussian
approximation. Lee and Kosterlitz~\cite{lee91} have pointed out
the inadequacy of the Gaussian approximation to produce shifts in
the locations of the energies of the equal-height peaks in
agreement with those described by Eqs.~(\ref{eq:10}) and
(\ref{eq:11}), which are rather well observed in simulations (see
Ref.~\cite{lee91} and Fig.~\ref{fig:2}(b)). In fact the
shortcomings of the Gaussian behavior has been critically
discussed in the early work of Challa et al.~\cite{challa86}. Our
analysis is therefore suggesting again the need for a more
realistic theory. It is tempting to assume that the attempted here
and well obeyed expansions for the energy cumulants, starting with
the correction term $bL^{-1}$, may have their origin in the
neglected domain wall corrections and could have the same
explanation with the existing puzzling situation concerning the
shift behavior of the free energy minima pointed out by Lee and
Kosterlitz~\cite{lee91}.

\begin{ack}
This research was supported by the Special Account for Research
Grants of the University of Athens under Grant Nos. 70/4/4071.
N.G. Fytas would like to thank the the Alexander S. Onassis Public
Benefit Foundation for financial support.
\end{ack}

{}

\newpage


\begin{table*}
\caption{\label{tab:1}Exact results for the bulk energies $e_{o}$
and $e_{d}$, the latent heat $\Delta e$~\cite{baxter73,kihara54},
and the values of $V_{4}(\infty)|_{min}$ and
$U_{2}(\infty)|_{max}$~\cite{lee91,janke93} for the $q=10$, $q=8$,
$q=6$, and $q=5$ Potts model in two dimensions. The last row
refers to the $R=1$ triangular SAF model.}
\begin{tabular}{cccccc}
\hline \hline Model & $e_{o}$ & $e_{d}$ & $\Delta e$ & $V_{4}(\infty)|_{min}$ & $U_{2}(\infty)|_{max}$ \\
\hline
$q=10$ Potts & -1.66425 & -0.96820 & 0.696 & 0.5589 & 1.0751 \\
\hline
$q=8$ Potts & -1.59673 & -1.11037& 0.486 & 0.6207 & 1.0333 \\
\hline
$q=6$ Potts & -1.50875 & -1.30775 & 0.201 & 0.6598 & 1.0051 \\
\hline
$q=5$ Potts & -1.47367 & -1.42075 & 0.053 &0.6662 & 1.0003 \\
\hline
Tr. SAF & -1.61784(728) & -1.48852(565) & 0.129(12) & 0.66435(33) & 1.00174(25)\\
\hline
\end{tabular}
\end{table*}

\end{document}